\documentclass[english,aps,prl,twocolumn,groupedaddress,nofootinbib]{revtex4-1}
\usepackage[LGR,T1]{fontenc}
\usepackage[latin9]{inputenc}
\usepackage{textcomp}
\usepackage{amssymb}
\usepackage{graphicx}
\usepackage{subscript}
\usepackage{amsmath}
\usepackage{color}
\usepackage{ulem}
\makeatletter

\DeclareFontEncoding{LGR}{}{}
\DeclareTextSymbol{\~}{LGR}{126}
\makeatother
\usepackage{babel}

\newcommand{\Tc}{T_{\rm c}}
\newcommand{\Tstar}{T^\star}
\newcommand{\pstar}{p^\star}
\newcommand{\Hc}{H_{\rm c2}}

\newcommand{\Ce}{C_{\rm e}}
\newcommand{\TBCO}{Tl$_{2}$Ba$_{2}$CuO$_{6+\delta}$}
\newcommand{\NdLSCO}{Nd$_{0.4}$La$_{1.6-x}$Sr$_{x}$CuO$_{4}$}
\newcommand{\EuLSCO}{Eu$_{0.2}$La$_{1.8-x}$Sr$_{x}$CuO$_{4}$}
\newcommand{\BSCO}{Bi$_{2+y}$Sr$_{2-x-y}$La$_x$CuO$_{6+\delta}$}
\newcommand{\LSCO}{La$_{2-x}$Sr$_{x}$CuO$_{4}$}
\newcommand{\YBCO}{YBa$_{2}$Cu$_{3}$O$_{y}$}
\newcommand{\mJmolK}{mJmol$^{-1}$K$^{-2}$}

\begin{document} 
\title{Normal state specific heat in the cuprates La$_{2-x}$Sr$_x$CuO$_4$ and Bi$_{2+y}$Sr$_{2-x-y}$La$_x$CuO$_{6+\delta}$ near the critical point of the pseudogap phase}

\author{C.~Girod$^{1,2}$, D.~LeBoeuf$^3$, A.~Demuer$^3$, G.~Seyfarth$^3$, S.~Imajo$^4$, K.~Kindo$^4$, Y.~Kohama$^4$, M.~Lizaire$^2$, A.~Legros$^2$, A.~Gourgout$^2$, H.~Takagi$^5$,
T.~Kurosawa$^6$, M.~Oda$^6$,  N.~Momono$^7$, J.~Chang$^8$, S.~Ono$^9$, G.-q. Zheng$^{10,11}$}
\author{C.~Marcenat$^{12}$}\email{christophe.marcenat@cea.fr}
\author{L.~Taillefer$^{2,13}$}\email{louis.taillefer@usherbrooke.ca}
\author{T.~Klein$^{1}$}\email{thierry.klein@neel.cnrs.fr}

\date{\today}
\address{$^1$ Univ. Grenoble Alpes, CNRS, Grenoble INP, Institut N\'eel, F-38000 Grenoble, France}
\address{$^2$ Institut quantique,  D\'epartement de physique \& RQMP, Universit\'e de Sherbrooke, Sherbrooke, Qu\'ebec J1K 2R1, Canada}
\address{$^3$ Univ. Grenoble Alpes, INSA Toulouse, Universit\'e Toulouse Paul Sabatier, EMFL, CNRS, LNCMI, F-38000 Grenoble, France}
\address{$^4$ Institute for Solid State Physics, University of Tokyo, Kashiwa, Chiba 277-8581, Japan}
\address{$^5$ Department of Advanced Materials, University of Tokyo, Kashiwa 277-8561, Japan}
\address{$^6$ Department of Physics, Hokkaido University, Sapporo 060-0810, Japan}
\address{$^7$ Muroran Institute of Technology, Muroran 050-8585, Japan}
\address{$^8$ Department of Physics, University of Zurich, CH-8057 Zurich, Switzerland}
\address{$^9$ Central Research Institute of Electric Power Industry, Materials Science Research Laboratory, 2-6-1 Nagasaka, Yokosuka, Kanagawa, Japan}
\address{$^{10}$ Department of Physics, Okayama University, Okayama 700-8530, Japan}
\address{$^{11}$ Institute of Physics, Chinese Academy of Sciences, and Beijing National Laboratory for Condensed Matter Physics, Beijing 100190, China}
\address{$^{12}$ Univ. Grenoble Alpes, CEA, Grenoble INP, IRIG, PHELIQS, 38000 Grenoble, France}
\address{$^{13}$ Canadian Institute for Advanced Research, Toronto, Ontario M5G 1M1, Canada}

\date{\today}
\begin{abstract}
The specific heat $C$ of the cuprate superconductors La$_{2-x}$Sr$_x$CuO$_4$ and Bi$_{2+y}$Sr$_{2-x-y}$La$_x$CuO$_{6+\delta}$ was measured at low temperature (down to 0.5~K), for dopings $p$ close to $p^\star$, the critical doping for the onset of the pseudogap phase. A magnetic field up to 35~T was applied to suppress superconductivity, giving direct access to the normal state at low temperature, and enabling a determination of $C_e$, the electronic contribution to the normal-state specific heat, at $T \to 0$. In La$_{2-x}$Sr$_x$CuO$_4$ at $x=p = 0.22$, 0.24 and 0.25, $C_e / T = 15-16$~mJmol$^{-1}$K$^{-2}$ at $T = 2$~K, values that are twice as large as those measured at higher doping ($p > 0.3$) and lower doping ($p < 0.15$). This confirms the presence of a broad peak in the doping dependence of $C_e$ at $p^\star\simeq 0.19$, as previously reported for samples in which superconductivity was destroyed by Zn impurities. Moreover, at those three dopings, we find a logarithmic growth as $T \to 0$, such that $C_e / T \sim {\rm B}\ln(T_0/T)$. The peak vs $p$ and the logarithmic dependence vs $T$ are the two typical thermodynamic signatures of quantum criticality. In the very different cuprate Bi$_{2+y}$Sr$_{2-x-y}$La$_x$CuO$_{6+\delta}$, we again find that $C_e / T \sim {\rm B}\ln(T_0/T$) at $p \simeq p^\star$, strong evidence that this $\ln(1/T)$ dependence -- first discovered in the cuprates La$_{1.8-x}$Eu$_{0.2}$Sr$_x$CuO$_4$ and La$_{1.6-x}$Nd$_{0.4}$Sr$_x$CuO$_4$ -- is a universal property of the pseudogap critical point. All four materials display similar values of the B coefficient, indicating that they all belong to the same universality class.
\\

\end{abstract}

\maketitle

\section{Introduction}
 
Unravelling the mystery of high temperature superconductivity remains a fundamental issue in modern solid state physics. A central question is the nature of
the enigmatic pseudogap phase which appears below a temperature $\Tstar$ and a critical hole concentration (doping) $\pstar$. Well above $\pstar$, the Fermi surface consists of a large quasi-2D cylinder (see for instance \cite{ARPES,ADMR} in \TBCO), the measured carrier concentration is equal to $n_{\rm H}=1+p$ (per CuO$_2$ plane) and the  Sommerfeld coefficient is on the order of 5~\mJmolK \cite{Bangura,Michon,Nakamae,Wang}. On the other hand, for $p\leq \pstar$, ARPES studies show that the Fermi surface breaks into small nodal  'Fermi arcs'  \cite{ARCS} and Hall effect measurements then indicate that the carrier concentration drops to $n_{\rm H}=p$ in \YBCO\ (YBCO) \cite{Badoux},  \NdLSCO\ (Nd-LSCO) \cite{Collignon} and  \BSCO\ (Bi2201) \cite{Putzke,Lizaire}. 

Specific heat measurements in the normal state of Nd-LSCO and \EuLSCO\ (Eu-LSCO) \cite{Michon} recently showed that the electronic contribution to the specific heat, $\Ce/T$, actually displays a  pronounced peak as function of doping at $p\sim \pstar$. Moreover, for $p$ close $\pstar$, the electronic specific heat displays a logarithmic temperature dependence~: $\Ce/T =$ B$ \ln(T_0/T)$ \cite{Michon}. Both behaviors are  typical thermodynamic signatures of quantum criticality. It is important to investigate whether those characteristic features are also present - or not - in other cuprates. We hence report here a study of the temperature and doping dependence of the electronic specific heat $\Ce$ in \LSCO\ (LSCO) and Bi2201 single crystals.

In LSCO, large  $\Ce/T$ values (on the order of $15$~\mJmolK\ at 2~K, are measured in the vicinity of the onset of the pseudogap phase. However, in contrast to previous measurements in Nd/Eu-LSCO \cite{Michon}, $\Ce/T$ remains large over an extended doping range, confirming the former indication for the presence of a {\it broad} maximum in the doping dependence of $\Ce/T$ observed in Zn substituted samples \cite{Momono}. Despite the  presence of a large (hyperfine) Schottky contribution, we also show that a $\ln(1/T)$ contribution has to be introduced in order to fit the temperature dependence of $\Ce/T$. Similarly, a clear $\ln(1/T)$ contribution to $\Ce/T$  and concomitant large $\Ce/T$ values (on the order of $13$~\mJmolK\ at 0.65~K) are observed in BSCO for $p \sim \pstar$, hence confirming the universality of those features. Very similar values of the B coefficient are observed in all compounds (close to $\pstar$) and the influence of the magnetic field on $T_0$ will be discussed.

 \begin{table*}
\caption{\label{table1} Sample name, critical temperature $\Tc$, chemical substitution rates ($x$ and $y$), estimated $T=0$ upper critical field ($^*$ from \cite{Wang08,Frachet}), mass, and B   coefficient in the B$\ln(T_0/T$) contribution to the specific heat of the Bi$_{2+y}$Sr$_{2-x-y}$La$_x$CuO$_{6+\delta}$  and La$_{2-x}$Sr$_{x}$CuO$_{4}$ single crystals.  Bi2201 samples \#1 and \#4 are the same as OD10K and OD18K in \cite{Lizaire}, respectively. Bi2201 samples \#2 and \#3 are the same as those for which the NMR Knight shift was reported in \cite{Kawasaki2020}, yielding estimates of the pseudogap temperature $\Tstar$ as plotted in Fig. 6. Last line : data from \cite{Michon}.}
\begin{ruledtabular}
\begin{tabular}{ccccccc}
name&$\Tc$ (K)&$x$&$y$&$\Hc(0)$ (T)&mass (mg)&B  (\mJmolK)\\
\hline
Bi2201\#1 & $\sim 11.5$ & 0 & 0.05 & $\sim 15$& 0.42&$2.8\pm0.5$\\
Bi2201\#2 & $\sim 10$ & 0.04 & 0 & $\sim 15$& 1.20&$3.2\pm0.5$\\
Bi2201\#3 & $\sim 13.5$ & 0.08 & 0 & $\sim 17$& 0.37&$2.5\pm0.5$ \\
Bi2201\#4 & $\sim 18$ & 0.20 & 0 & $\sim 22$ (at 2.1~K)& 0.39&$< 1$\\
\hline
LSCO\#1 & 0 & 0 & - & 0 & 0.60&0\\
LSCO\#2 & 0 & 0.04 & - & 0 & 0.95&0\\
LSCO\#3 & $\sim 20$ & 0.12 & - & $\sim 19$& 0.75&0\\
LSCO\#4 & $\sim 25$ & 0.145 & - & $40\pm5^*$& 1.50&0\\
LSCO\#5 & $\sim 25$ & 0.22 & - & $37\pm4^*$& 0.94& $2.0\pm 0.3$\\
LSCO\#6 & $\sim 16$ & 0.24 & - & $\sim 23$ (at 2.1~K)& 1.50& $2.2\pm 0.3$\\
LSCO\#7 & $\sim 15$ & 0.25 & - & $\sim 14$ (at 2.1~K)& 0.85&$2.1 \pm 0.3$\\
\hline
Eu-LSCO & 11&$p=0.24$&&11& - &$2.5 \pm 0.5$\\
\end{tabular} 
\end{ruledtabular}
\end{table*}

\begin{figure}
\includegraphics[width=8cm]{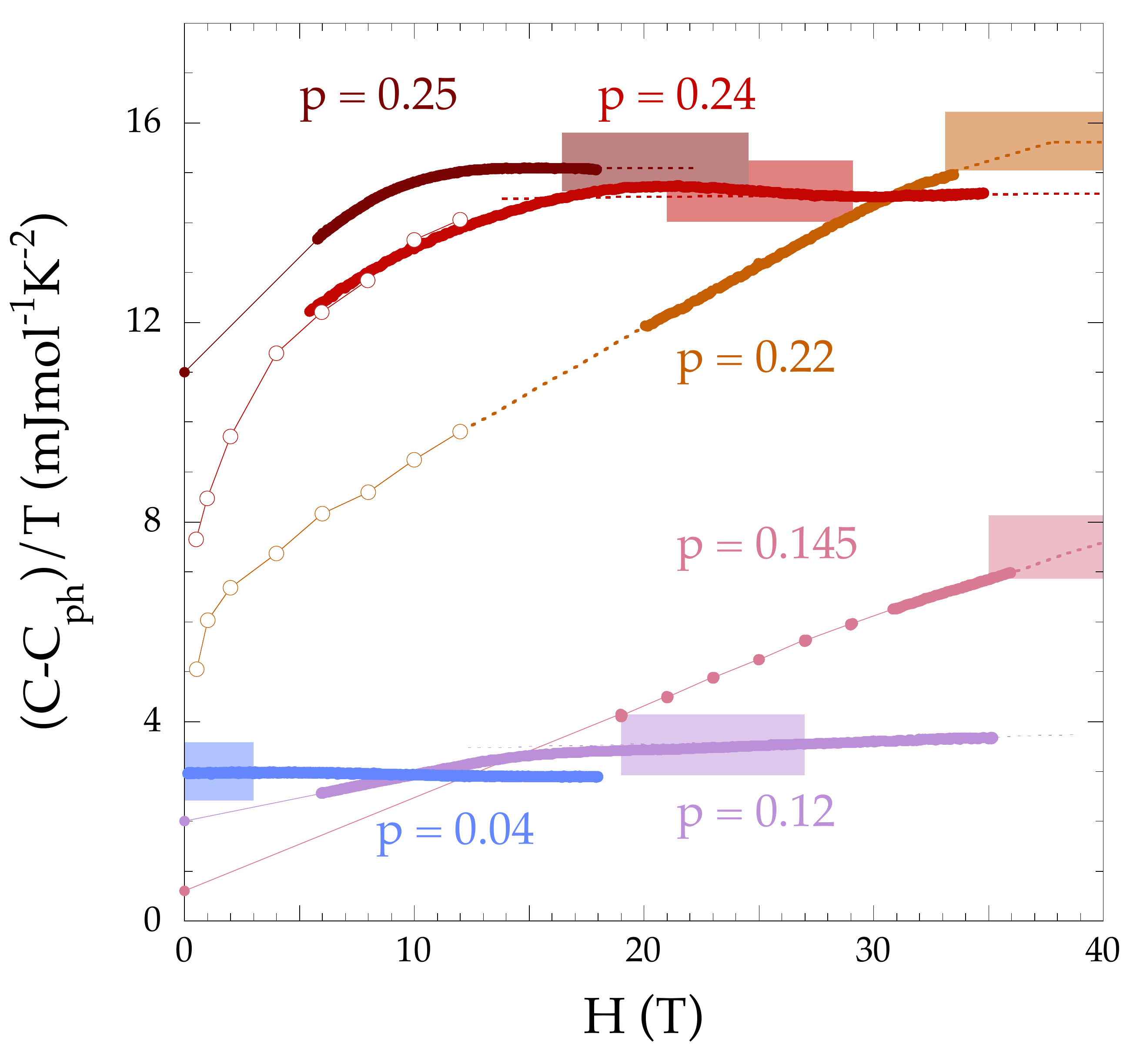}
\caption{Magnetic field dependence of the specific heat at $T \sim 2.1$ K in \LSCO\ (sample details are shown in Tab.~1), after subtraction of the phonon contribution $C_{\rm ph}$ (see Fig.~3). The small overshoot observed for $p\sim 0.24$  is most probably reminiscent of the superconducting transition in presence of strong fluctuations. The shaded (colored) boxes indicate the locus of the previously estimated $\Hc(0)$ values \cite{Wang08,Frachet}. Open circles are low field data extracted from \cite{Wang}. As seen, $(C-C_{\rm ph})/T$ increases with $p$, reaching $\sim 15-16$~\mJmolK\ in the normal state  for $p\sim 0.22-0.25$ at $T \sim$  2~K (see also Fig.~2).}
\end{figure}

\section{Methods}

The specific heat of seven LSCO single crystals with $0\leq p \leq 0.25$ and four Bi2201 single crystals with doping contents close to the onset of the pseudogap phase has been measured by AC micro-calorimetry down to $\approx 0.5$~K and up to 35~T. Sample LSCO\#3 and LSCO\#4 are the same samples than measured in references \cite{Frachet,Chang1}. A transport study of sample  Bi2201\#1 and Bi2201\#4 (labelled OD10K and OD18K, respectively) has recently been performed in \cite{Lizaire}.  The temperature $T^*$ was determined by the temperature dependence of the Knight shift  for Bi2201 \#2 and \#3 \cite{Kawasaki2020}

The heat capacity has been measured by a modulation technique where a periodically modulated heating power $P_{\rm ac}$ is applied at a frequency 
$2 \omega$. Recording the amplitude of the induced temperature oscillations $|T_{\rm ac}|$ and its thermal phase shift $\phi$ relative to the power allows to calculate the heat capacity : $C_p = P_{\rm ac} $sin$(-\phi) / 2\omega |T_{\rm ac}|$. A miniature CERNOX resistive chip has been split into two parts and attached to a small copper ring with PtW($7~\%$) wires. The first half ($R_H$) was then used as an electrical eater ($P_{\rm ac} = R_Hi_{\rm ac}^2(\omega)/2$) and the second ($R_T$) was used to record the temperature oscillations ($V_{ac}(2\omega)=({\rm d}R_T/{\rm d}T)T_{\rm ac}(2\omega)i_{\rm DC}$, where $i_{\rm DC}$ is a DC reading current). In order to subtract the heat capacity of the addenda (chip + a few $\mu$g of Apiezon grease used to glue the sample onto the back of the chip), the empty chip   was measured prior to the sample measurements. A precise \textit{in situ} calibration and corrections of the thermometers in magnetic field were included in the data treatment. This technique enabled us to obtain absolute values of the specific heat  of minute single crystals with an absolute accuracy better than $\sim 95~\%$ as deduced from measurements on ultra pure copper (for further details, see \cite{Michon}). 

Sample Bi2201\#2 has also been measured  in pulsed magnetic fields up to 39~T and down to 0.6~K by the heat pulse method described in reference \cite{Kohama}. The single crystal was mounted on the sample platform with a small amount of Apiezon grease and the total heat capacity was estimated by $C_p = Q/\Delta T$, where $Q$ is the applied heat and $\Delta T$ the resultant temperature change. The addenda contribution originating from the sample platform and  Apiezon grease was estimated by a separate experiment. The thermometer was calibrated by isothermal measurements of the magnetoresistance and the accuracy of $C_p$  was estimated to $\sim 93\%$ below  2~K and $\sim 90\%$ between 2 and 4~K, as deduced from measurements on a reference sample of polycristaline Ge \cite{Kohama}. 

\section{Results and discussion}
\subsection{La$_{2-x}$Sr$_{x}$CuO$_{4}$}

Efforts to measure the normal state specific heat at low temperature in LSCO have been hindered by the very large upper critical field values exceeding 60~T around optimal doping (for $0.14 \leq p \leq 0.21$  \cite{Wang08}, see also \cite{Frachet} and references therein). This implies that measurements in the normal state  have been limited to highly underdoped or overdoped samples in which superconductivity is absent \cite{Wang,Komiya,Nakamae} or weak enough to be easily suppressed by moderate magnetic fields \cite{Wang,Wang08}.  However, ARPES measurements indicated that the pseudogap critical point $0.17 \leq \pstar \leq 0.22$ \cite{Chang2,Yoshida}. Accordingly, resistivity measurements suggest that $\pstar = 0.19\pm 0.02$ \cite{Cooper,Laliberte,Boebinger}, \textit{i.e.} lying in the doping range for which the normal state is unattainable
 
\begin{figure}
\includegraphics[width= 8cm]{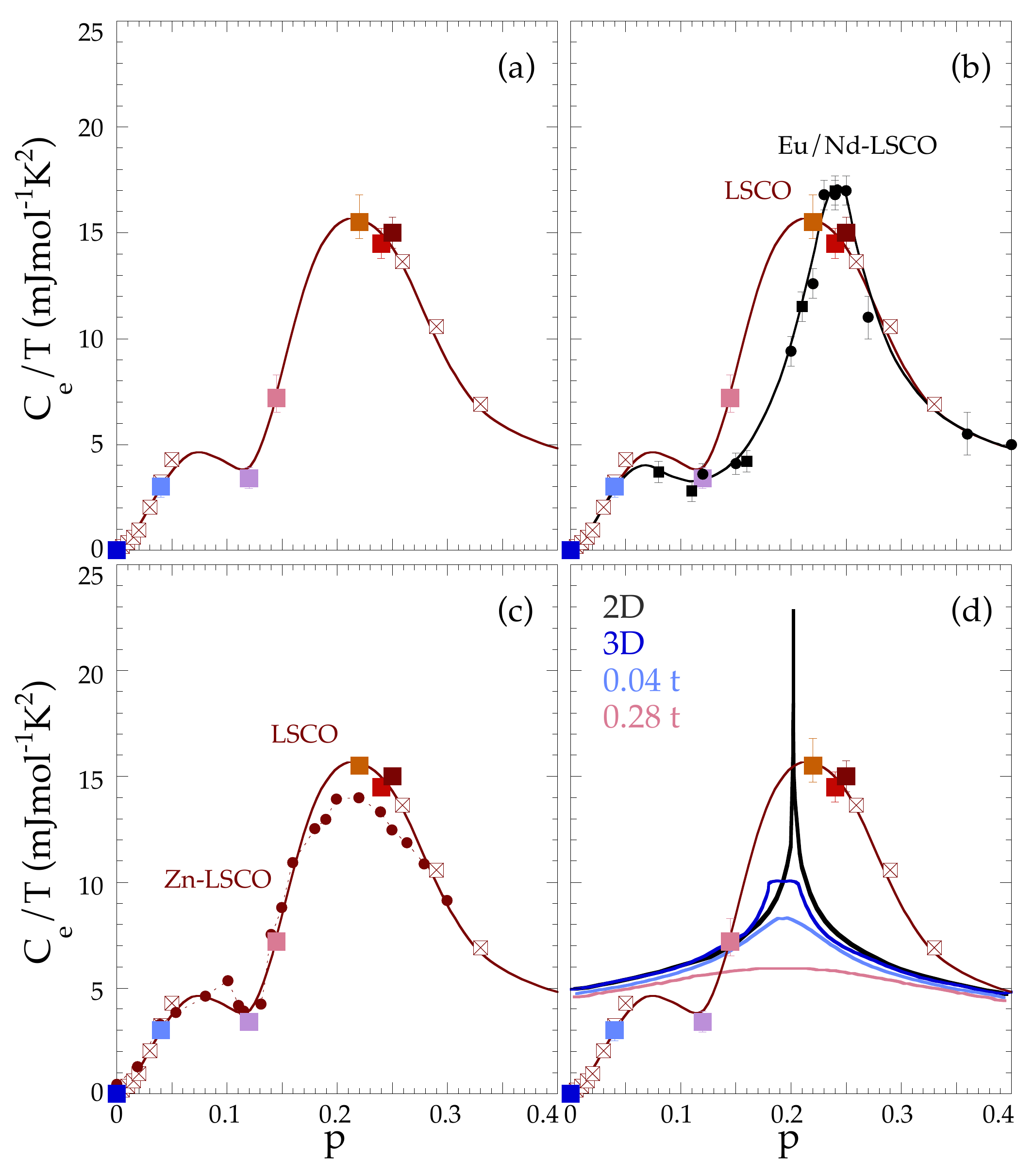}
\caption{(a) Electronic contribution to the specific heat $\Ce/T$ (at 2~K) as a function of the doping content $p$ in La$_{2-x}$Sr$_{x}$CuO$_{4}$ (solid squares, see Tab.~1 for sample details) together with data previously obtained in samples of lower $\Hc$ values \cite{Wang,Komiya,Nakamae} (crossed open squares). The same guide to the eyes (thin lines) is used in all panels. (b) same data as panel (a) together with data previously obtained in Nd-LSCO (black circles) as well as Eu-LSCO (black squares) crystals (from \cite{Michon}). (c) same data as in panel (a) together with data previously obtained in samples in which superconductivity was destroyed by Zn impurities \cite{Momono} ((red) closed circles). (d)  same data as in panel (a) together with calculations for a van Hove singularity at $p_{\rm vHs}=0.2$ (from \cite{Horio}) (solid lines) for the indicated geometries (2D/3D) and scattering rates ($\hbar/\tau$, see \cite{Horio} for details). As shown, the dependence obtained for any van Hove singularity only very poorly reproduces the experimental data. }
\end{figure}

\begin{figure}
\includegraphics[width=7.5cm]{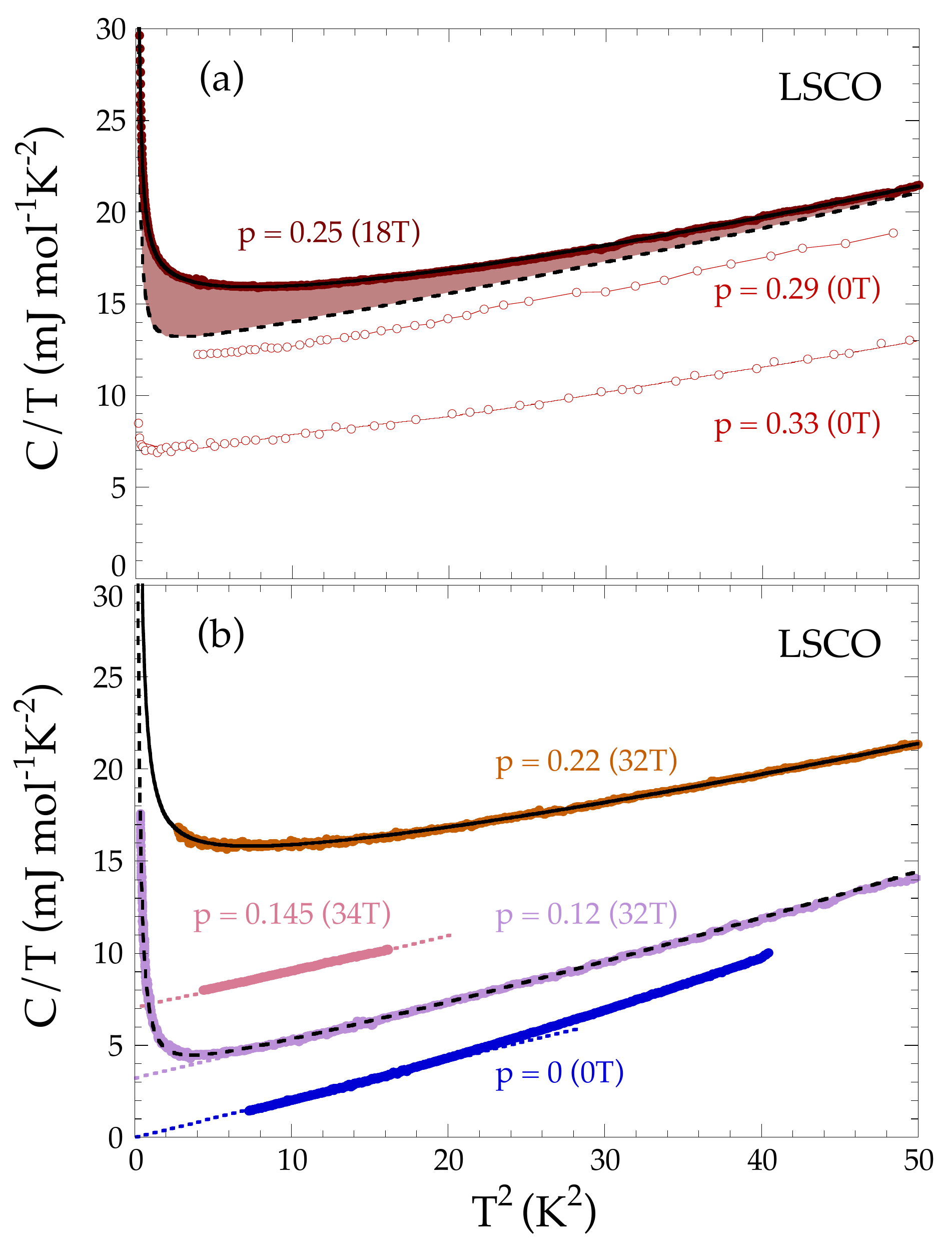}
\caption{Temperature dependence of the specific heat obtained in La$_{2-x}$Sr$_{x}$CuO$_{4}$ for $p \geq 0.25$ (panel a) and $p\leq 0.22$ (panel b)  (see Tab.~1 for sample details). For $p=$ 0.22 and 0.25, the data can be very well described by a $C/T = {\rm A}H^2/T^3 + \beta T^2 +\delta T^4+{\rm B}\ln(T_0/T)$ law ((black) solid lines). The (black) dashed line in panel (a) corresponds to the dependence that would be obtained for B $=0$ ($p = 0.25$) and the shaded area then highlights the contribution of the B$\ln(T_0/T)$ term. On the other hand a very good with to the data can be obtained with B $=0$ for $p=0.12$ ((black) dashed line in panel (b)). The standard $C/T = \gamma_0 + \beta T^2$ dependence is indicated by the dotted lines (panel b). The hyperfine (${\rm A}H^2/T^3$) contribution becomes fully negligible above $2-3$~K (see for instance $p=0.12$). The data at $p=0.29$ and $p=0.33$ (open symbols) are from \cite{Wang} and \cite{Nakamae}, respectively (see also supplemental materials for other compositions). }
\end{figure}

By applying magnetic fields up to 35~T, we were able to determine the specific heat in  the normal state  for $p$ values as close as possible to $\pstar$. Indeed, as shown in Fig.~1, the specific heat increases with field in the mixed state and saturates at high fields, clearly indicating that the applied fields are large enough to suppress superconductivity (or are slightly below $\Hc(0)$ for $p\sim$ 0.22 and 0.145). Note that a $H^2$ field dependence is observed at the lowest temperatures for $H\geq \Hc(0)$, indicating the presence of a hyperfine Schottky contribution : $C_{\rm hyp} = {\rm A}H^2/T^2$ with ${\rm A}\sim 4 \pm 1 \times 10^{-3}$~\mJmolK (see supplemental materials and Fig.~3 below). This hyperfine term is unfortunately hindering any unambiguous determination of the electronic contribution at very low temperature and we have here limited our lowest temperature to $\sim 2$~K for the analysis of $\Ce/T$ as $C_{\rm hyp}/T$ becomes negligible above this temperature  (see Fig.4 in supplemental materials for the relative contribution of each term).

Our measurements then clearly indicate that the normal state electronic contribution to the specific heat displays a maximum upon doping (see Fig.~2~(a)), similar to the one observed in Nd and Eu substituted LSCO \cite{Michon} (Fig.~2~(b)). As shown, very similar $\Ce/T$ values are obtained for $p\geq 0.25$, but clear differences are visible for lower dopings. Indeed, the position of the peak is shifted towards lower dopings in LSCO compared to its Nd and Eu substituted counterparts, due to the lower $\pstar$ value.  However, the maximum is significantly broader in LSCO, in agreement with  the $p$ dependence of the electronic specific heat previously obtained in Zn substituted samples \cite{Momono} (see Fig.~2~(c)). Even though their exists some uncertainty on the determination of $p$, it is worth noting that those large $\Ce/T$ values have been observed in samples with upper critical fields ranging from $\Hc\sim 14$~T to $\Hc \geq 34$~T (see Fig.~1) unambiguously confirming that $\Ce/T$ remains large over an extended doping range (from $p \approx 0.20$ to $p \approx 0.26$).

Finally, note that attempts to determine the electronic contribution from high temperature specific heat measurements (\textit{i.e.} for $T>\Tc$) led to unconclusive results for $p = 0.20$. Those determinations were based on a difficult phonon subtraction procedure and reported values vary from 7 \cite{Loram89} to 14~\mJmolK\ \cite{Matsuzaki}. The latter value seems to be consistent with our work (see Fig.~2), but $\Ce/T$ actually decreases with $T$, hindering any direct comparison between those high $T$ values and those obtained here for $T \rightarrow 0$. 

The change of topology of the Fermi surface at $p_{\rm vHs}\gtrsim \pstar$ is expected to give rise to a strong (diverging) van Hove singularity in a purely 2D system \cite{Horio} (see black line in Fig.~2~(d)). In fact, the amplitude of this singularity is strongly reduced both by the dimensionality and  the presence of disorder. The dark-blue line in Fig.~2~(d) is a calculation for an anisotropic 3D system with an transverse hopping rate $t_z=0.07 t$, $t$ being the nearest-neighbour hopping parameter in the tight binding Hamiltonian. The light-blue and purple lines are calculations for scattering rates $\tau$ equal to : $\hbar/\tau = 0.04t$ and $0.28t$, respectively \cite{Horio} (the latter value is consistent with the measured residual resistivity $\rho_0 \sim 30$ $\mu\Omega$cm). We deduce from Fig.~2~(d) that the computations poorly reproduce the experimental data. Moreover, as disorder is very different in pristine LSCO and Zn substituted samples, one should have observed a much reduced peak in the latter. This is not the case, hence strongly suggesting that the observed peak is not related to this van Hove singularity.

Finally, it is interesting to compare our data on LSCO at $p = 0.04$ with recent quantum oscillation measurements in the 5-layer cuprate Ba$_2$Ca$_5$Cu$_{10}$(F,O)$_2$~\cite{Kunisada2020}. The frequency of the oscillations coming from the innermost layer (labelled IP0), in which there is long-range antiferromagnetic order, is $F = 147$~T.  ARPES measurements show that the Fermi surface in this metal with antiferromagnetic order consists of four small closed hole pockets at nodal locations in the Brillouin zone, centered approximately at $(\pm \pi/2, \pm \pi/2)$~\cite{Kunisada2020}. The 2D carrier density $n$ contained by each hole pocket is given by $n = F / \Phi_0$, where $\Phi_0 = h / 2e$. The hole concentration (doping) $p$ should then be given by $p = 2n$, given that there are two hole pockets per magnetic Brillouin zone, which yields $p \simeq 0.04$. The effective mass extracted from the quantum oscillations is $m^* \sim 0.7m_e$, where $m_e$ is the bare electron mass. This corresponds to a specific heat Sommerfeld coefficient $\gamma = 2 \times 1.43  \times  m^* \sim 2$~\mJmolK~mol-Cu, where the factor 2 comes from having two hole pockets in the Brillouin zone. This expected value of $\gamma$ is not far from the value of 3~\mJmolK~mol-Cu reported here for LSCO at $p = 0.04$~(Fig.~1 and 2). Given that the carrier density estimated from the Hall number in LSCO at $p = 0.04$ is indeed $n = n_{\rm H} \simeq 0.04$~\cite{Ando2004-LSCO-Hall},  the Fermi surface of LSCO at $p = 0.04$ should be the same as that observed in the inner plane of Ba$_2$Ca$_5$Cu$_{10}$(F,O)$_2$, {\it i.e.} four nodal hole pockets containing $p$ carriers, even though the commensurate antiferromagnetic phase in LSCO ends at $p \simeq 0.02$. This points to a similar Fermi surface in the pseudogap phase, beyond $p \simeq 0.02$, and in the antiferromagnetic phase, as suggested by Hall studies in various cuprates~\cite{Badoux,Collignon,Lizaire} and associated calculations~\cite{Storey,Eberlein2016,Verret2017,Chatterjee2017}. 
 
A second indication for the presence of a QCP is the occurence of a $\ln(1/T)$ contribution in the temperature dependence of the specific heat \cite{Michon}. As shown in Fig.~3, the data can be well described by the standard $C/T = \gamma_0  + \beta T^2$ law above $\sim 2$~K for dopings far from the critical doping $\pstar$, \textit{i.e.} for  $p\leq 0.145$ and $p\geq 0.29$ (a small $\delta T^4$ correction has to be introduced above $\sim 5$~K).
At very low temperature, a clear deviation due to the hyperfine contribution is visible, in good agreement with the $H^2$ dependence observed in $C/T(H)$ (see Fig.2 in the supplemental materials).  In contrast with data at $p=0.12$, the upturn at low temperature cannot be  described by the hyperfine contribution alone for dopings close to $\pstar$ (\textit{i.e.} for $p = 0.22-0.25$), indicating an extra contribution which extends well above 2~K. Note that a very similar  contribution is also visible in the data previously obtained by Wang {\it al.} \cite{Wang} for $p \sim 0.26$ (see Fig.~1 in supplemental materials). Such an upturn has also been observed in Zn-LSCO (below $\sim 3.5$~K, for $x \sim 0.2$) and has been attributed to local magnetic moments induced by Zn doping \cite{Momono}, but this interpretation can not hold in our pristine LSCO crystals. 

In this doping range, the data can then be very well fitted by introducing an extra $\ln(1/T)$) contribution to the specific heat, $C/T = {\rm A}H^2/T^3 + \beta T^2 + \delta T^4 + {\rm B}\ln(T_0/T)$  where the constant B$\ln(T_0)$ term contains both the band  structure Sommerfeld contribution and a cutoff temperature above which the contribution of the (quantum) fluctuations vanishes (see solid lines in Fig.~3). To clearly highlight this logarithmic contribution, it is displayed  in Fig.~3~(a) as a shaded area under the curve for $p=0.25$. On the contrary, a very good fit to the data can be obtained with $ {\rm B}=0$ for $p=0.12$ (see dashed line in Fig.~3~(b)).

\begin{figure}
\includegraphics[width=8.5cm]{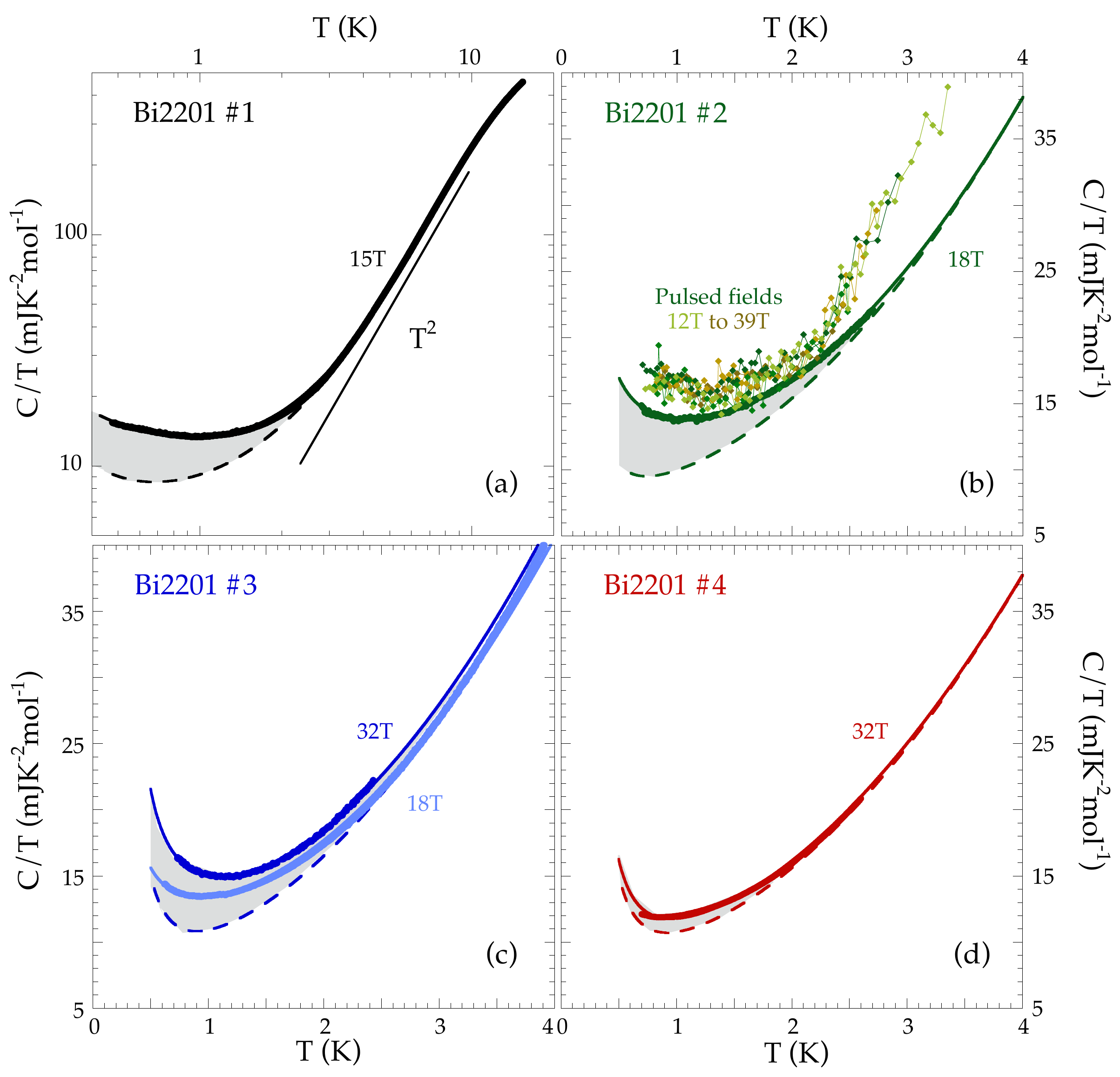}
\caption{Temperature dependence of the specific heat of our \BSCO\ single crystals (see Tab.~1 for sample details). Solid lines are fit to the data assuming that $C/T = {\rm A}H^2/T^3 + \beta T^2 + {\rm B}\ln(T_0/T)$ and the dashed line correspond to the dependence that would be obtained for $B=0$. The contribution of the $ {\rm B}\ln(T_0/T)$ term is highlighted by the grey shaded area. As shown panel (d), the contribution of the $\ln(1/T)$ term is much weaker in Bi2201\#4 of larger $\Tc$ value. Complementary pulsed field data for fields ranging from 12~T to 39~T (Bi2201\#2) are also displayed in panel (b). Measurements of Bi2201 at both 18~T and 32~T are displayed  in panel (c), indicating a field dependence of $C/T$ in the normal state, well above 1~K (see also Fig.~6 and text for details). }
\end{figure}

\subsection{Bi$_{2+y}$Sr$_{2-x-y}$La$_x$CuO$_{6+\delta}$ }

We performed similar measurements on four Bi2201 crystals (see Tab.~1). In Bi2201, the hyperfine contribution to the specific heat we observe is much weaker than in LSCO ($ {\rm A}=8 \pm 1 \times 10^{-4}$~\mJmolK) and the normal state could be reached in all measured samples (see Table~1 and Fig.3 in the supplemental materials for the upper critical field $\Hc(0)$ values). As an example, in the normal state of sample Bi2201\#1, at a field just above $\Hc(0) \sim 15$~T, the  hyperfine term is about 20 times smaller than in LSCO. The contribution of the $\ln(1/T)$ term can then be straightforwardly identified in the temperature dependence of $C/T$ and is highlighted by a shaded area in Fig.~4~(a) (see Fig.4 in supplemental materials for the relative contribution of each term). 

Complementary measurements have also been performed in Bi2201\#2 in pulsed magnetic fields up to $\sim 39$~T (see Fig.~4~(b)). Although slightly larger, the obtained $C/T$ values are in reasonable agreement with the one obtained in DC field. They confirm the presence of a clear upturn at low temperature and  the absence of any significant contribution from the hyperfine term (\textit{i.e.} no field dependence) up to 39~T on the whole temperature range.
In all samples, $C/T$ can then be very well fitted by a ${\rm A}H^2/T^3 + \beta T^2 + {\rm B}\ln(T_0/T)$ law (solid lines in Fig.~4). In samples Bi2201\#1-3, the amplitude of the logarithmic contribution is identical within error bars  (see Table~1) and this contribution is only significantly smaller in Bi2201\#4, which has the highest $\Tc$ and hence a doping content further away from $\pstar$ (see shaded area in Fig.~4~(d)). 

As pointed out by Lohneysen in the context of heavy fermion systems \cite{Lohneisen}, the universality of the critical exponents of a given universality class is expected to translate into a constant B value for all compounds belonging to this universality class. It is then important to note that we obtained very similar B coefficients ($\sim 2-3$~\mJmolK\  close to $\pstar$) in LSCO, Nd/Eu-LSCO and Bi2201 (see table 1), strongly indicating that they all belong to the same universality class. Moreover, following Varma \cite{Varma}, B is expected to be on the order of the band structure Sommerfeld coefficient $\gamma_0$ and our B value is hence {\it quantitatively} consistent with a $\gamma_0$ value being on the order of 5~\mJmolK\, as measured far from $\pstar$ (the coupling constant $g=B/\gamma_0 \sim 0.5$ is then in agreement with transport and photoemission measurements, see \cite{Varma} for a detailed discussion). 

On the other hand, B$\ln(T_0)$ varies from $\sim 11-12$ ~\mJmolK\ in Bi2201 to $\sim 16-19$~\mJmolK\  in lanthanum based cuprates (see Fig.5). This increase in $T_0$ is consistent with Seebeck coefficient measurements \cite{Lizaire,Daou} showing that, at low temperature, $S(T)$ is significantly larger in Nd-LSCO than in Bi2201. $S(T)$ also displays a $\ln(1/T)$ temperature dependence but this contribution is then flattening off above $\sim 30$~K in Bi2201 whereas it extends up to $\sim 100$~K in Nd-LSCO.  Finally, note that we observe a small increase of $T_0$ with field in Bi2201\#3 (see Fig.~4~(c) and Fig.~5). Although small, this field induced change can not be attributed to the Schottky contribution (see Fig.5 in supplemental materials) and still has to be understood (see also concluding remarks below).

Finally let us turn to the doping dependence of $\Ce/T$ in Bi2201. It is difficult to determine an unambiguous doping content $p$ in Bi$_{2+y}$Sr$_{2-x-y}$La$_x$CuO$_{6+\delta}$ due to the various doping routes ($x$, $y$ and $\delta$) and the phase diagram has hence been plotted as a function of the critical temperature $\Tc$ in Fig.~6.  As shown, the electronic specific heat is on the order of $8 \pm 1$~\mJmolK\ at 3~K but rapidly increases with decreasing $T$, reaching $\sim 13 \pm 1$~\mJmolK\ at 0.65~K in Bi2201\#1-3. As shown, $\Ce/T$ (at 0.65~K) then decreases in Bi2201\#4. Bi2201\#1-3 clearly lie close to the onset of the pseudogap phase as deduced from ARPES  \cite{Kondo} and NMR \cite{Kawasaki} studies, emphasising the enhancement of $\Ce/T$ (at low $T$) in the vicinity of this point. Note that a linear temperature dependence of the resistivity has been reported recently in Bi2201\#1 (labelled OD10K) \cite{Lizaire} but this linearity does not persist down to $T = 0$~K, indicating that the doping content is slightly below $\pstar$ for this $T_c$ value. 

It is interesting to further compare our specific heat data on Bi2201 with recent transport measurements performed on same samples (sample labelled OD10K and OD18K in ref.~\cite{Lizaire} are the same as Bi2201\#1 and Bi2201\#4, respectively). In particular, the Hall effect measurements yield a large difference in Hall number between the two, namely $n_{\rm H} \sim 1.4$ for Bi2201\#1 and $n_{\rm H} \sim 0.75$ for Bi2201\#4, interpreted as a sharp, nearly two-fold, drop in carrier density with decreasing $p$~\cite{Lizaire}.  A similar decrease has been reported by Putzke {\it et al.} \cite{Putzke} and a very similar drop was observed previously in YBCO~\cite{Badoux} and Nd-LSCO~\cite{Collignon}. This drop has been identified as a key signature of the pseudogap phase that reveals a transformation of the Fermi surface across $\pstar$, consistent with a change from a large surface containing $1+p$ holes to small Fermi pockets containing $p$ holes~\cite{Proust2019,Storey}. Note that for the same Bi2201 samples, the electronic specific heat at $T = 3$~K changes very little from one doping to the other (Fig.~6). This is not necessarily incompatible with a transformation of the Fermi surface. Indeed, calculations show that when a metal undergoes a transition into a phase of antiferromagnetic order (at $T \simeq 0$)~\cite{Verret2017}, the specific heat is barely affected initially whereas the Hall number exhibits a rapid drop. 

Another property of Bi2201 that was measured in the field-induced normal state at $T \to 0$ is the NMR Knight shift ~\cite{Kawasaki}, which reflects the electronic spin susceptibility, typically assumed to be proportional to the  electronic density of states. In that sense, it should be closely related to the electronic specific heat reported here. What we observe is that the two quantities do not show the same degree of change below $\pstar$. Indeed, while the electronic spin susceptibility at $T \simeq 2-3$~K drops by a factor of about 20~\% in going from $\Tc = 10$~K to $\Tc = 18$~K~\cite{Kawasaki}, the electronic specific heat does not change (Fig.~6). This suggests that the two measurements, performed in the very same conditions of temperature and field, on the same Bi2201 samples, do not reflect the same underlying physical quantity, {\it i.e.} they do not both reflect the same density of states. Understanding this difference, revealed here for the first time in a cuprate material, should provide new insight into the nature of the pseudogap phase.

 \begin{figure}
\includegraphics[width=8.5cm]{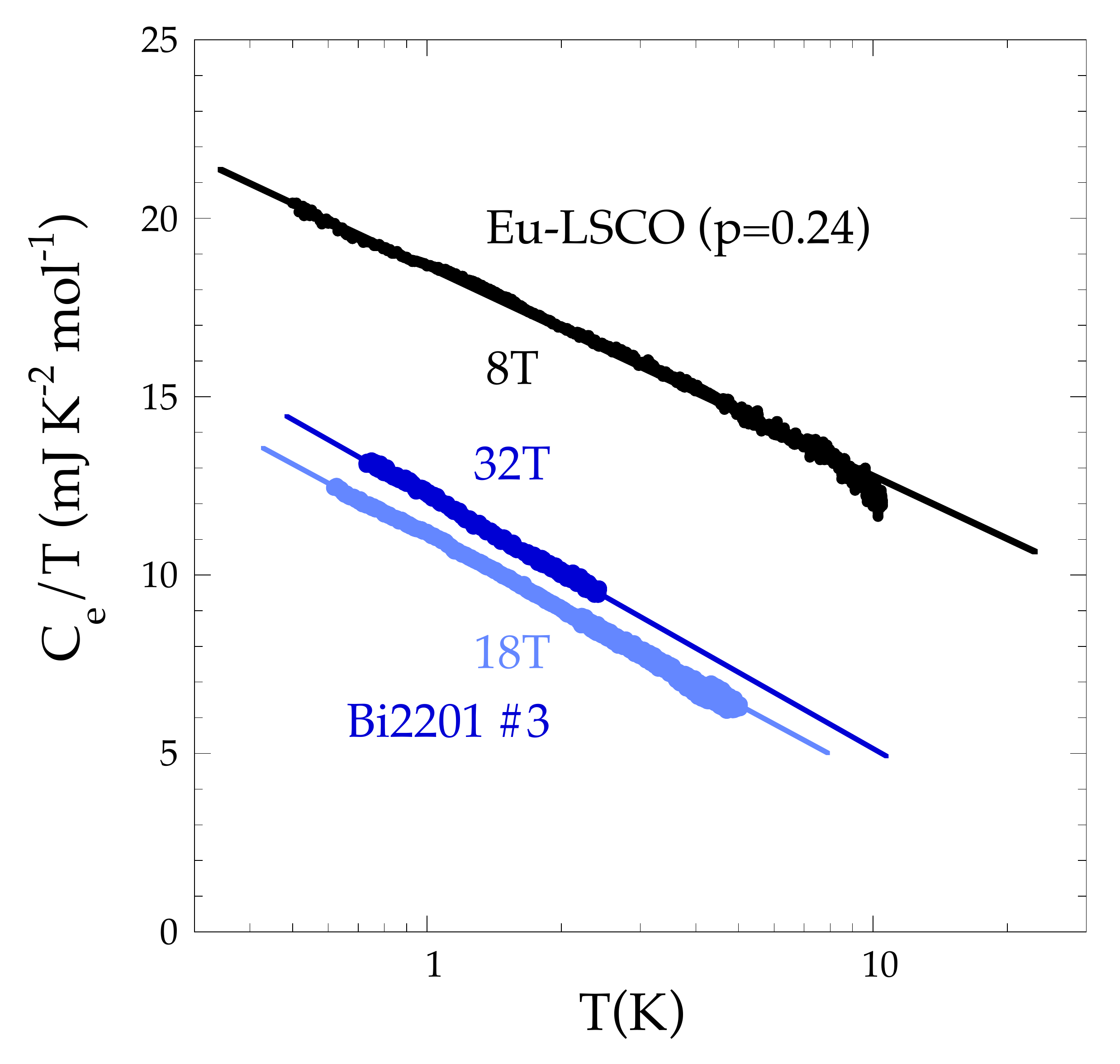}
\caption{Temperature dependence of the electronic contribution to the specific heat  $\Ce/T=C/T - $A$H^2/T^3 - C_{\rm ph}/T$ of Bi2201\#3, compared to the dependence previously obtained in Eu-LSCO (from \cite{Michon}), clearly displaying the logarithmic decrease ($\Ce/T={\rm B}\ln(T_0/T)$), with a field dependent $T_0(H)$.}
\end{figure}

\begin{figure}
\includegraphics[width=8.5cm]{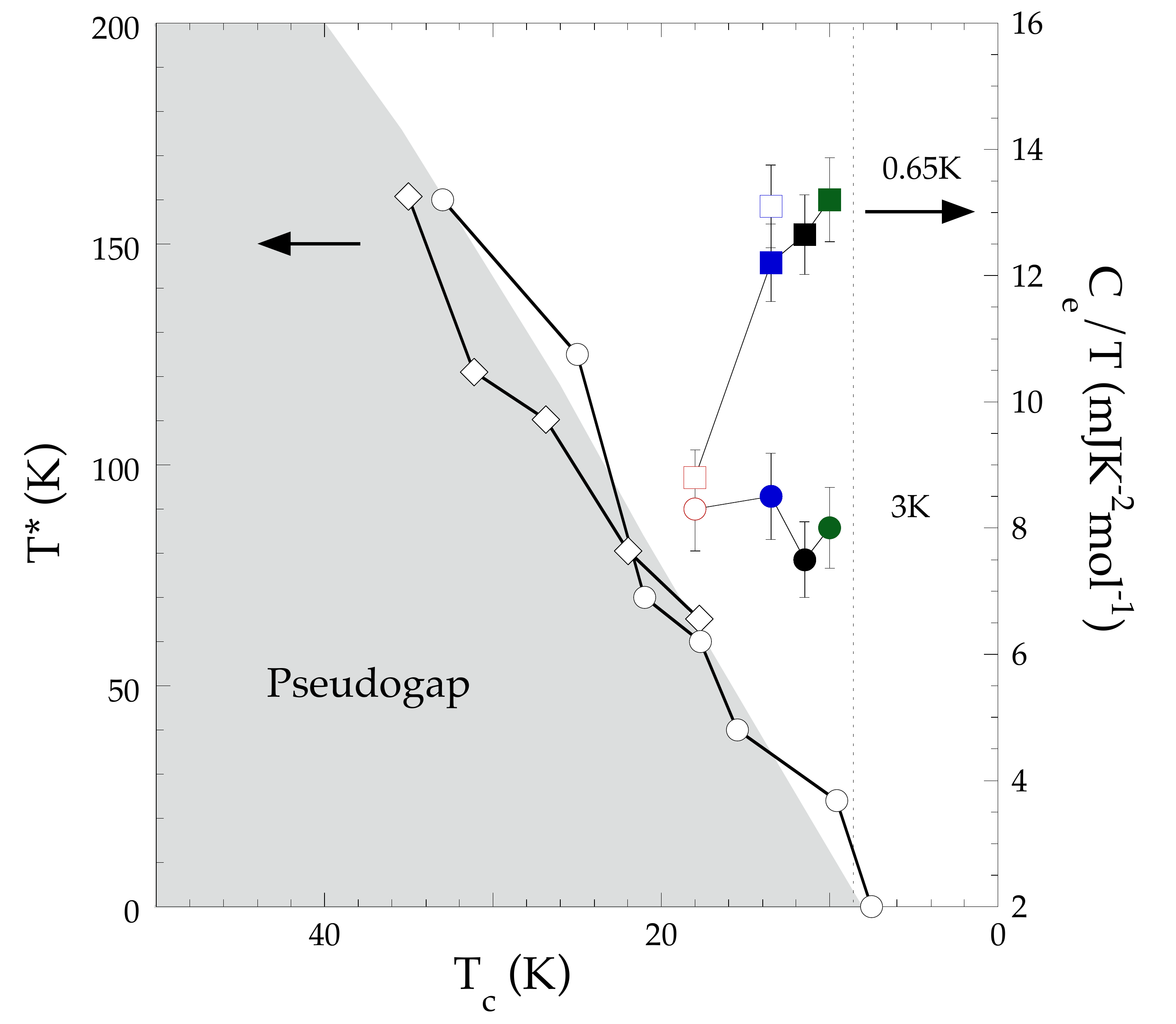}
\caption{Electronic specific heat $\Ce/T$  as a function of the critical temperature $\Tc$ at 0.65~K (squares) and 3~K (circles)  for Bi2201 samples \#1 (black), \#2 (green), \#3 (blue) and \#4 (red) (same color code as in Fig.4), together with the evolution of the pseudogap phase onset temperature $\Tstar$, as deduced from ARPES (diamonds from \cite{Kondo}) and NMR (circles from \cite{Kawasaki2020,Kawasaki}) measurements. Closed symbols correspond to 18~T measurements and open symbols to 32~T measurements, suggesting a (small) increase of the electronic contribution with field in the normal state (see Fig.~5).}
\end{figure}
 
\section{Concluding remarks}

In summary, we have  shown that the electronic contribution to the normal state specific heat displays a $\ln(1/T)$ temperature dependence associated to a strong increase of $\Ce/T$ for $T \rightarrow 0$, close to the critical doping  $\pstar$ that marks the onset of the pseudogap phase in both LSCO and Bi2201, as previously observed in Nd/Eu-LSCO \cite{Michon}. These features seem therefore generic in cuprates and are classical signatures of the existence of a QCP, whose  nature  has  not  yet been identified. In LSCO, the  $\ln(1/T)$ term is observed up to 0.26 {\it i.e.} well above $\pstar=0.19\pm0.02$. This extended doping range could be reminiscent of the {\it anomalous} form of criticality previously pointed out by Cooper {\it et al.} in transport measurements \cite{Cooper}, for which a a linear term in the temperature dependence of the resistivity is observed from $p \sim 0.18$ up to $p \sim 0.3$  \cite{Cooper}. 

Moreover, we observed that the criticality is reinforced by the magnetic field in Bi2201 (the large hyperfine contribution and $\Hc$ values are hindering a similar study in LSCO) as the electronic contribution to the specific heat slightly increases with field in the normal state. This puzzling behaviour is in contrast to the one usually observed in heavy fermion systems in the vicinity of a magnetic QCP \cite{QCP-heavy fermions}. Indeed, in this case the quantum fluctuations are suppressed by the application of magnetic field and the $\ln(1/T)$ contribution saturates at low $T$ for large $H$ values. Recent NMR and sound velocity measurements \cite{Frachet} indicated that the short-range magnetism is reinforced under high magnetic fields in LSCO. The interplay between this magnetic state and the  signatures of quantum criticality observed here still has to be clarified, but it is worth noting that this magnetic phase vanishes at a critical doping which extrapolates to $\pstar$ for  large magnetic field leading to a possible field dependence of the QCP (see also \cite{Sachdev}). However, no indication for the presence of such a magnetic state in the vicinity of the onset of the pseudogap phase has been observed in Bi2201.

Finally, calculations  \cite{Sordi1} for the two-dimensional Hubbard model in the doped Mott insulator regime have proposed an alternative interpretation of these thermodynamic anomalies without invoking broken symmetries. A sharp cross-over in the specific heat can occur  when crossing the Widom line emanating from a finite temperature critical endpoint of a first-order transition between a pseudogap phase  and a "standard" metal with Fermi liquid correlations \cite{Sordi2}. 

The work in Grenoble was supported by the Laboratoire d'excellence LANEF (ANR-10-LABX-51-01) and was performed at the LNCMI, a member of the European Magnetic Field Laboratory (EMFL). Work at the LNCMI was supported by the French Agence Nationale de la Recherche (ANR) (contract ANR-19-CE30-0019). L.T. acknowledges support from the Canadian Institute for Advanced Research (CIFAR) and funding from the Natural Sciences and Engineering Research Council of Canada (NSERC; PIN:123817), the Fonds de recherche du Quebec Nature et Technologies (FRQNT), the Canada Foundation for Innovation (CFI), and a Canada Research Chair. This research was undertaken thanks in part to funding from the Canada First Research Excellence Fund. Part of this work was funded by the Gordon and Betty Moore Foundation's EPiQS Initiative (Grant GBMF5306 to L.T). J.C. acknowledge support from the Swiss National Science Foundation).

\end{document}